\documentclass[10pt,aps,prl,twocolumn,preprintnumbers,nofootinbib]{revtex4-2}
\usepackage{amsmath,amssymb,mathtools}
\usepackage{amsfonts}
\usepackage{hyperref}
\usepackage{xspace}
\usepackage{graphicx}
\usepackage{grffile}
\usepackage{enumitem}
\usepackage{xcolor}
\usepackage{booktabs}
\newcommand{\dd}{\mathrm{d}}

\newcommand{\GeV}{\mathrm{GeV}}
\newcommand{\TeV}{\mathrm{TeV}}
\newcommand{\code}[1]{\texttt{#1}}
\newcommand{\NNLOJET}{\textsc{NNLOjet}\xspace}

\newcommand{\NPCnlo}{\code{NPC23\_NLO}\xspace}
\newcommand{\BDSS}{\code{BDSS23\_NLO}\xspace}
\newcommand{\NNFF}{\code{NNFF11\_NLO}\xspace}
\newcommand{\HAPS}{\code{HAPS\_NNLO}\xspace}
\newcommand{\PDF}{\code{PDF4LHC21}\xspace}

\begin{document}

\preprint{CERN-TH-2026-238, ZU-TH 36/26}

\title{\boldmath Precise QCD Predictions for Photon-Associated Hadron Production at the LHC}

\author{Gabriele Fiore$^a$, Aude Gehrmann-De Ridder$^{a,b}$, Alexander Huss$^c$,\\ Francesco Merlotti$^a$, Giovanni Stagnitto$^d$}
\affiliation{$^a$ Institute for Theoretical Physics, ETH, CH-8093 Zürich, Switzerland\\
             $^b$ Physik-Institut, Universit\"at Z\"urich, Winterthurerstrasse 190, 8057 Z\"urich, Switzerland\\
             $^c$ Theoretical Physics Department, CERN, 1211 Geneva 23, Switzerland\\
             $^d$ Universit\`{a} di Genova \& INFN, Via Dodecaneso 33, I-16146 Genova, Italy}

\begin{abstract}
Prompt-photon production with recoiling QCD radiation is crucial for new-physics searches,
determinations of parton distribution and fragmentation functions, and jet-quenching studies in heavy-ion collisions.
We present the first fully differential NNLO QCD predictions for photon-associated identified-hadron production at the LHC.
We compare predictions obtained with different fragmentation function (FF) sets to ALICE and ATLAS data at $\sqrt{s}=5.02\,\TeV$.
Our calculation provides an important theoretical ingredient for FF global fits
and offers a precision proton--proton baseline for photon-tagged heavy-ion studies.
\end{abstract}

\maketitle

\textit{Introduction---}\ignorespaces
Highly energetic photons provide a clear experimental signature in the final state of collider events.
Prompt-photon production constitutes a background to new-physics searches~\cite{Lindert:2017olm} and probes
the gluon distribution in the proton~\cite{Carminati:2013lvm,Campbell:2018wfu}.
The accompanying hadronic final state provides access to parton fragmentation, both inclusively and through
hadrons identified inside reconstructed jets~\cite{ATLAS:2019dsv}.
It further enables precision QCD studies, including energy correlators~\cite{Generet:2026vto}.
In heavy-ion collisions, the photon also provides a calibrated reference for jet-quenching studies: being essentially unaffected by the quark--gluon plasma,
it constrains the transverse momentum of the recoiling parton prior to energy loss~\cite{Casalderrey-Solana:2015vaa,Kang:2017xnc}.

The perturbative description of prompt-photon production has reached a high level of accuracy over the past decade.
State-of-the-art calculations include NNLO QCD predictions for isolated-photon and photon-plus-jet
production~\cite{Campbell:2016lzl,Campbell:2017dqk,Chen:2019zmr}, with realistic photon-isolation criteria~\cite{Chen:2022gpk},
as well as isolated-photon plus two-jet production~\cite{Badger:2023mgf}.
Electroweak effects in prompt-photon production have been investigated, initially through purely weak corrections~\cite{Maina:2004rb,Kuhn:2005gv},
and later at full NLO EW accuracy for $\gamma$ + jet~\cite{Lindert:2017olm}.
Higher-order EW Sudakov effects, which are enhanced in the high-energy tails of distributions, have been studied at two-loop order~\cite{Kuhn:2005gv,Becher:2013zua,Becher:2015yea}.
Combined EW and QCD predictions for isolated-photon and photon-plus-jet production were discussed in Refs.~\cite{Schwartz:2016olw,Campbell:2017dqk,Campbell:2018wfu,Lindert:2017olm}.

In this Letter, we present the first fully differential NNLO QCD predictions for the production of a prompt photon and an identified charged hadron at the LHC.
We consider both the {\em semi-inclusive} process,
\begin{equation}\label{eq:proc1}
  p p \to \gamma + h + X\,,
\end{equation}
and the {\em hadron-in-jet} process,
\begin{equation}\label{eq:proc2}
  p p \to \gamma + j(h) + X\,.
\end{equation}

Both processes are relevant for the extraction of fragmentation functions (FFs),
which encode non-perturbative information related to parton-to-hadron transitions.
FFs are routinely determined by global fits to single-inclusive $e^+e^-$ annihilation (SIA),
semi-inclusive deep-inelastic scattering (SIDIS), and hadron-collider data.
Most global FF analyses have been performed at NLO~\cite{Bertone:2018ecm,Borsa:2021ran,Khalek:2021gxf,Moffat:2021dji,Gao:2024dbv}.
NNLO determinations have been obtained from SIA data~\cite{Anderle:2015lqa,Bertone:2017tyb,Soleymaninia:2018uiv,Abdolmaleki:2021yjf} and subsequently extended to combined SIA and SIDIS data, first with approximate NNLO SIDIS corrections~\cite{Borsa:2022vvp,AbdulKhalek:2022laj} and more recently with the complete coefficient functions~\cite{Gao:2025hlm}.
NNLO coefficient functions are available for SIA~\cite{Rijken:1996ns,Mitov:2006wy} and SIDIS~\cite{Goyal:2023zdi,Bonino:2024qbh,Goyal:2024emo}.
Perturbative predictions for these processes have recently been extended to N$^3$LO~\cite{He:2025hin,Dong:2026uvw}.
At hadron colliders, NNLO QCD predictions have recently become available for single-inclusive identified-hadron production~\cite{Czakon:2025yti}.

Photon--hadron production and its sensitivity to fragmentation functions have previously been studied at NLO in QCD~\cite{deFlorian:2010vy,Ochoa-Oregon:2023ktx}.
Our work thus provides a necessary theoretical ingredient towards the
inclusion of these photon-plus-hadron datasets in a global NNLO FF fit.
In particular, hadron-collider data and hadron-in-jet observables offer important sensitivity to the gluon FF~\cite{Gao:2024nkz,Bonino:2026dvr}.
Additionally, precise proton--proton predictions provide the baseline needed to disentangle nuclear effects on fragmentation in heavy-ion collisions~\cite{CMS:2018mqn,ALICE:2026sue}.

In the following, we first briefly present our methodology and then compare
our predictions to LHC data at
$\sqrt{s} = 5.02\,\TeV$ from the ALICE~\cite{ALICE:2020atx} and
ATLAS~\cite{ATLAS:2019dsv} Collaborations.
\newline

\textit{Methodology---}\ignorespaces
The fully differential cross section for the processes in Eqs.~\eqref{eq:proc1}--\eqref{eq:proc2} reads
\begin{align}\label{eq:xs}
  \dd\sigma^{\gamma+h} = \sum_{i,j,p} & \int \dd x_1 \dd x_2 \dd z \, \mathcal{L}_{ij}(x_1,x_2,\mu_f^2) D_p^h(z,\mu_a^2) \nonumber \\
  & \times \dd \hat{\sigma}_{ij,p}(x_1, x_2, z,\mu_r^2,\mu_f^2,\mu_a^2) \, ,
\end{align}
where $\mathcal{L}_{ij}$ is the incoming parton luminosity, while the fragmentation function $D_p^h$
describes the fragmentation of the parton $p$ into the identified
hadron $h$ carrying a momentum fraction $z$ of the parent parton.
Collinear factorisation is performed at the factorisation scale
$\mu_f$ for $\mathcal{L}_{ij}$ and at the fragmentation scale $\mu_a$ for $D_p^h$.
The short-distance partonic cross section admits an expansion in the renormalised strong coupling constant $\alpha_s$:
\begin{align}
  \dd \hat{\sigma}_{ij,p} = \dd \hat{\sigma}^{\mathrm{LO}}_{ij,p}+\left(\frac{\alpha_s}{2\pi}\right) \dd \hat{\sigma}^{\mathrm{NLO}}_{ij,p}+\left(\frac{\alpha_s}{2\pi}\right)^2 \dd \hat{\sigma}^{\mathrm{NNLO}}_{ij,p}+\ldots \, .
\end{align}
Beyond leading order, real and virtual contributions to the short-distance
cross section produce infrared divergences, which need to be regulated in the
intermediate steps of the calculation.
With the introduction of subtraction terms, the infrared divergences associated
with unresolved configurations in real radiation matrix elements are removed, such
that the remainder is suitable for numerical integration.
The subtraction terms, after analytical integration, are added back at the
virtual level and combined with mass factorisation contributions.
We use the antenna subtraction
method~\cite{Gehrmann-DeRidder:2005btv,Currie:2013vh} for fully differential NNLO calculations.

As for initial-state partons, the identification of a parton in the final state
spoils the complete cancellation of collinear divergences, since resolving an
individual parton within a collinear cluster introduces collinear sensitivity.
The antenna subtraction method has been extended
in~\cite{Gehrmann:2022cih,Gehrmann:2022pzd,Bonino:2024adk} to deal with the
additional final-state collinear configurations.
Other NNLO subtraction methods, such as the sector-improved residue subtraction scheme~\cite{Czakon:2010td}, are also capable of producing predictions for identified-hadron processes~\cite{Czakon:2021ohs,Generet:2025bqx}.

In this work, we implement photon-associated identified-hadron production at
hadron colliders through NNLO accuracy within the parton-level event generator
\NNLOJET~\cite{NNLOJET:2025rno}.
The subtraction terms for this process are new within the
antenna subtraction formalism: this is the first process with
identified hadrons at hadron colliders computed at NNLO accuracy in this framework.
Previous applications include predictions for hadron-in-jet observables in association
with a $Z$ boson at the LHC at NLO accuracy~\cite{Caletti:2024xaw} and, more
recently, hadron-in-jet observables in $e^+e^-$ annihilation at NNLO accuracy~\cite{Bonino:2026dvr}.

We perform the standard \NNLOJET checks on our implementation~\cite{NNLOJET:2025rno}.
Moreover, we validate our NLO predictions for the isolated-photon plus hadron
process against the DIPHOX code~\cite{Binoth:2002wa}, finding excellent agreement.

We employ the $G_\mu$ scheme to compute the electroweak parameters with the Fermi
constant $G_F = 1.1663787 \times 10^{-5}\,\GeV^{-2}$, and the $Z$- and
$W$-boson masses $M_Z = 91.1876\,\GeV$ and $M_W = 80.379\,\GeV$, respectively.
The NNLO PDF set employed to produce all predictions is \PDF~\cite{PDF4LHCWorkingGroup:2022cjn}.
The reference value of the strong coupling constant is $\alpha_s(M_Z)=0.118$,
and its evolution is performed at three loops with LHAPDF6~\cite{Buckley:2014ana}.

We consider theoretical predictions obtained with four different FF sets: \BDSS~\cite{Borsa:2023zxk}, \NNFF~\cite{Bertone:2018ecm}, \NPCnlo~\cite{Gao:2024dbv}, and \HAPS~\cite{Soleymaninia:2026xjq}.
The \BDSS and \NPCnlo FF sets include SIA, SIDIS, and $pp$ data in their global fits.
In particular, the ATLAS data analysed in this work
are included in the \NPCnlo fit.
The dataset of the \NNFF global analysis consists of SIA and $pp$ data.
Lastly, the \HAPS FF set is fitted to SIA and SIDIS data.
We choose \HAPS as the default FF set to match the highest perturbative accuracy of the parton-level cross section.

For the normalised distributions considered below, the choice of renormalisation $\mu_r$, factorisation $\mu_f$, and fragmentation $\mu_a$ scales deserves particular attention and is specified separately for each process below.
Higher-order theoretical uncertainties are estimated with a 15-point scale variation of $\mu_r$, $\mu_f$, and $\mu_a$ around the central scales for the numerator and a 7-point scale variation of $\mu_r$ and $\mu_f$ for the denominator.
In the ratio, we treat the scales as uncorrelated, i.e., $\mu_r$ and $\mu_f$ are varied independently between numerator and denominator, while we restrict $1/2 \leq \xi'/\xi \leq 2$ for any pair of scales, with $\xi$ denoting the multiplicative variation with respect to the corresponding central scale.
We always match the perturbative accuracy of the denominator to that of the numerator.

In order to separate direct photons produced in the hard scattering from the photons resulting from fragmentation processes and hadronic decays, experimental measurements usually impose an isolation prescription to limit the hadronic activity in the vicinity of the photon of interest.
To approximate the experimental isolation requirements, we use hybrid photon isolation~\cite{Siegert:2016bre}, combining smooth-cone isolation~\cite{Frixione:1998jh} with parameters $R_{\rm in}=0.1$, $\epsilon=0.1$ and $n=1$, and a fixed-cone requirement that the hadronic transverse energy within $R_{\rm iso}>R_{\rm in}$ remains below $E_T^{\rm max}$.
The inner smooth cone suppresses photon--parton collinear configurations, avoiding the need for parton-to-photon fragmentation functions, while the outer-cone parameters $R_{\rm iso}$ and $E_T^{\rm max}$ follow the experimental selections and are specified below for each dataset.
\newline

\textit{Comparison to ALICE Data---}\ignorespaces
The ALICE Collaboration measured the photon-associated semi-inclusive production of identified light charged hadrons $h^\pm$, based on data collected during LHC runs at $\sqrt{s} = 5.02\,\TeV$~\cite{ALICE:2020atx}.
The measurement is differential in the observable
\begin{align}\label{eq:x}
  z_T = \frac{p_T^h}{p_T^\gamma} \,,
\end{align}
where $p_T^h$ is the transverse momentum of the identified hadron, and $p_T^\gamma$ is the transverse momentum of the isolated photon.
Note that at leading order, $z_T \equiv z$, the fragmentation momentum fraction.
Differential cross sections are normalised to the fiducial cross section $\sigma^\gamma$ for isolated-photon production.

Isolated photons are required to have transverse momentum $p_T^\gamma \in [12, 40]\,\GeV$ and absolute pseudorapidity $|\eta^\gamma| < 0.67$.
The fixed-cone isolation parameters are $R_{\rm iso} = 0.4$ and $E_{T}^{\rm max} = 1.5\,\GeV$.
Identified charged hadrons are required to have transverse momentum $p_T^h \in [0.5,10]\,\GeV$ and absolute pseudorapidity $|\eta^h| < 0.8$.
Additional cuts on the absolute azimuthal angle difference $|\Delta \varphi| \in [7\pi/8,\pi]$ and the pseudorapidity distance $|\Delta\eta| \leq 1.2$ between the isolated photon and the identified hadron are used to select the events entering the analysis.

For the scales, we adopt $\mu_r = \mu_f = p_T^\gamma$. For semi-inclusive observables, the most natural choice for the fragmentation scale is to set it equal to the renormalisation and factorisation scales, hence we employ $\mu_a = p_T^\gamma$.

\begin{figure}[ht]
  \centering
  \includegraphics[width=\columnwidth]{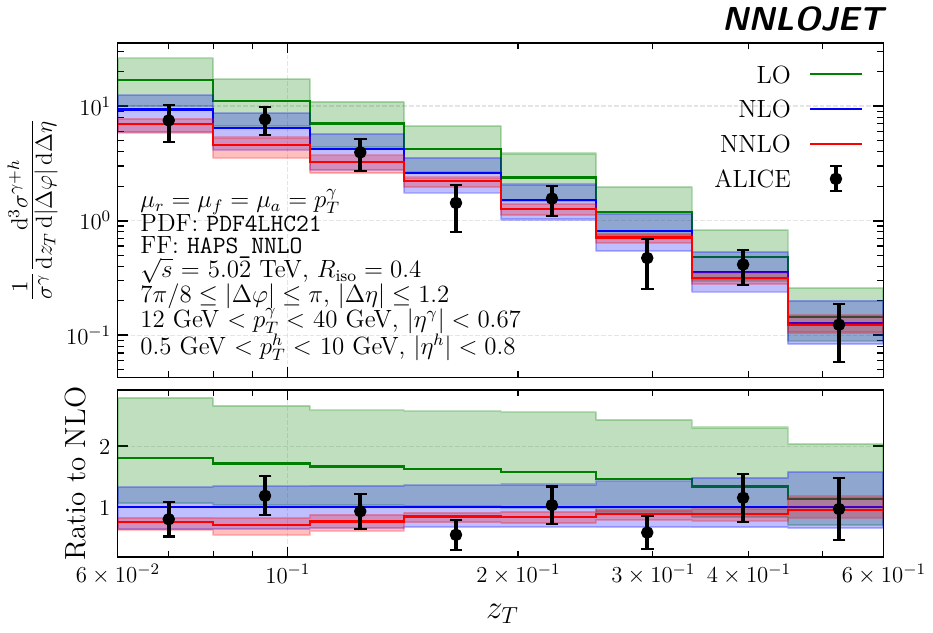}
  \caption{Data--theory comparison for semi-inclusive $h^\pm$ production in association with an isolated photon up to NNLO with \HAPS.}
  \label{fig:alice-absolute}
\end{figure}
In Fig.~\ref{fig:alice-absolute} we compare our theoretical predictions at different perturbative orders to ALICE data.
The agreement with data improves substantially when NLO corrections are included.
Given the large experimental uncertainties, the NNLO corrections do not yet visibly improve the description of the data.
Nevertheless, we observe excellent perturbative convergence when the $\mathcal{O}(\alpha_s^3)$ corrections are taken into consideration, and a reduction of scale-variation uncertainties.

\begin{figure}[ht]
  \centering
  \includegraphics[width=\columnwidth]{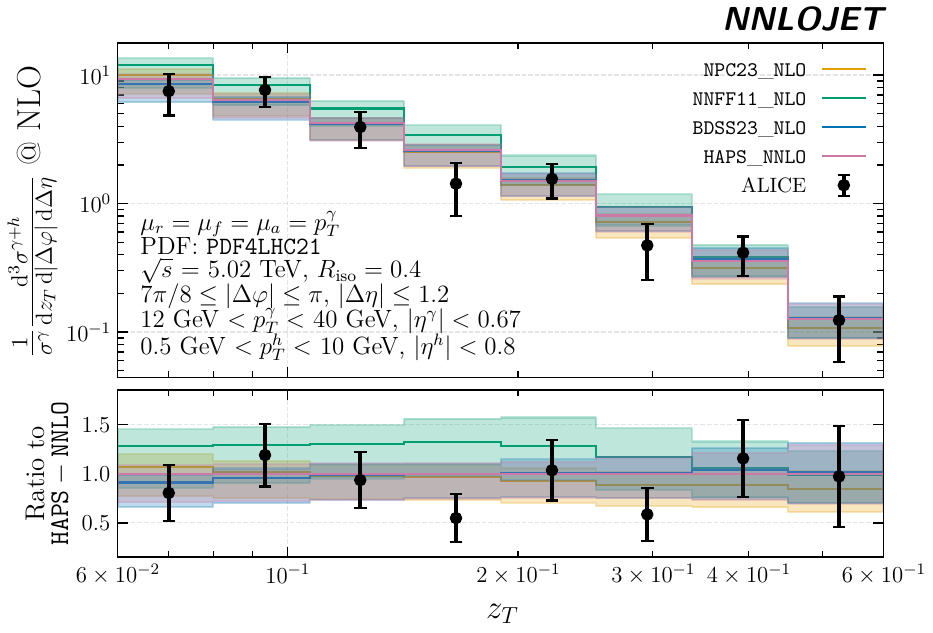}
  \caption{Data--theory comparison for semi-inclusive $h^\pm$ production in association with an isolated photon at NLO for the hard scattering and different FF sets.}
  \label{fig:alice-comparison}
\end{figure}
To assess the impact of different FF sets on theoretical predictions in the ALICE experimental setup, in Fig.~\ref{fig:alice-comparison} we compare predictions with different sets at NLO.
Overall, the four sets describe the data well within theoretical uncertainties across the whole $z_T$ range:
only the predictions computed with the \NNFF set lie slightly above those obtained with the other FF sets, which overlap almost perfectly across the full kinematic range.

\begin{figure*}[ht]
  \centering
  \includegraphics[width=\textwidth]{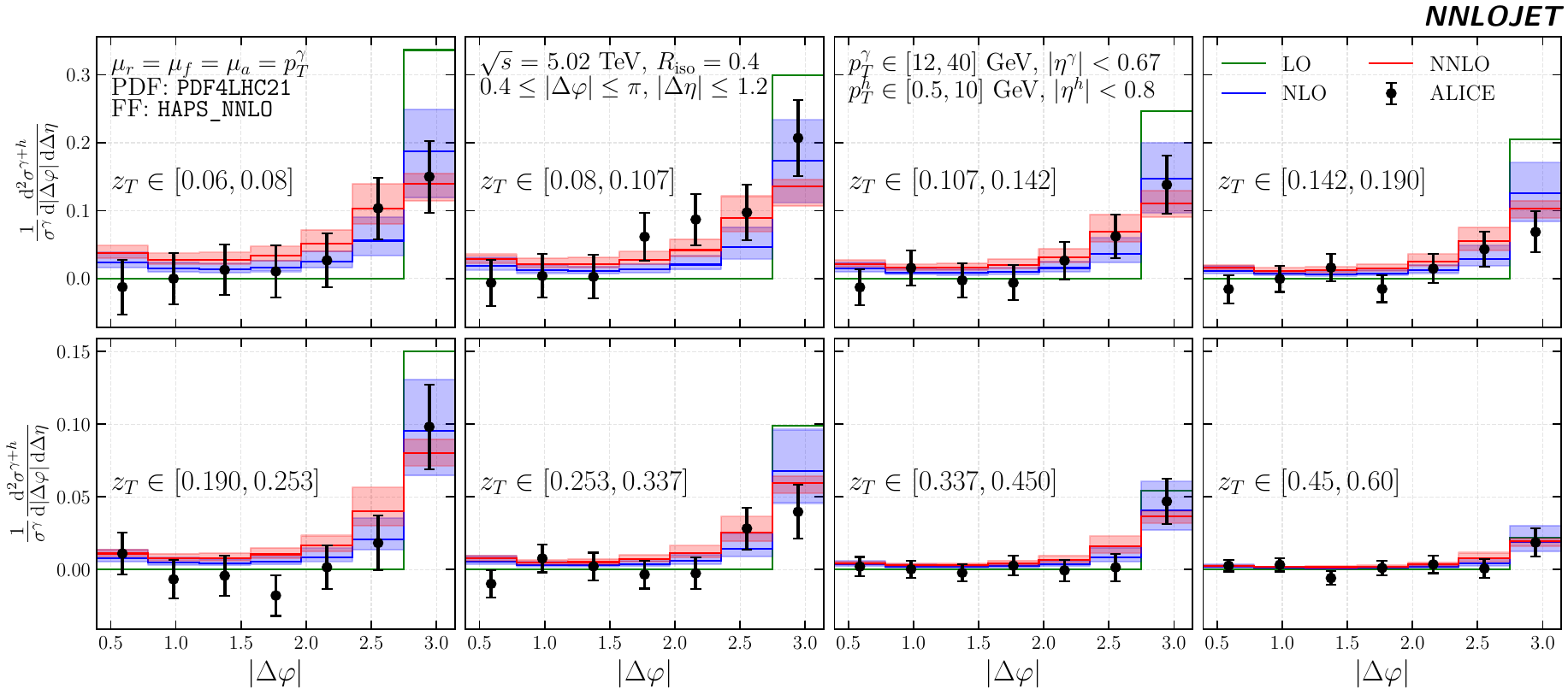}
  \caption{Data--theory comparison for $h^\pm$ production in association with an isolated photon up to NNLO with \HAPS. Data for the $z_T$ bins $[0.06, 0.08]$, $[0.142,0.190]$, and $[0.45,0.60]$ are provided in~\cite{ALICE:2020atx}. Data for the other five $z_T$ bins are provided in~\cite{Torales-Acosta:2021zwt}.}
  \label{fig:alice-ang-corr}
\end{figure*}
The cross section presented in Figs.~\ref{fig:alice-absolute} and \ref{fig:alice-comparison} is differential in $z_T$, and the $|\Delta \varphi| \in [7\pi/8,\pi]$ cut selects the away-side region containing the back-to-back LO configuration.
In Fig.~\ref{fig:alice-ang-corr} we present the semi-inclusive hadron production cross section differential in the absolute azimuthal separation $|\Delta \varphi|$ between the isolated photon and the identified hadron, in the same eight $z_T$ bins shown in Figs.~\ref{fig:alice-absolute} and \ref{fig:alice-comparison}.
Bins not containing $|\Delta\varphi|=\pi$ first receive contributions at
$\mathcal{O}(\alpha\alpha_s^2)$, one power of $\alpha_s$ beyond the Born process.
Our NNLO calculation therefore provides NLO accuracy in these bins,
while retaining NNLO accuracy in the rightmost bin containing the Born contribution.
The NNLO prediction improves the description of the data at large $|\Delta \varphi|$ values, although both the NLO and NNLO curves remain compatible with the data, given the large experimental uncertainties.
\newline

\textit{Comparison to ATLAS Data---}\ignorespaces
The ATLAS Collaboration has measured hadron-in-jet cross sections at $\sqrt{s} = 5.02\,\TeV$~\cite{ATLAS:2019dsv}.
The measurements are differential in the transverse momentum of the hadron $p_T^h$ and in
\begin{equation}\label{eq:zh}
  z^h = \frac{p_T^h}{p_T^{j(h)}} \cos \Delta R\,.
\end{equation}
In Eq.~\eqref{eq:zh} $p_T^{j(h)}$ is the transverse momentum of the jet containing the hadron, and $\Delta R$ is the angular distance in the pseudorapidity--azimuth plane between the identified hadron and the axis of the jet that encloses it.
As for $z_T$, at LO $z^h$ coincides with the fragmentation variable $z$.
The distributions are normalised to the inclusive jet rate $\sigma^{\gamma+j}$ for the selected photon--jet pairs.

Isolated photons are required to have transverse momentum $p_T^\gamma \in [80,126]\,\GeV$ and absolute pseudorapidity $|\eta^\gamma| < 2.37$, excluding the region $|\eta^\gamma| \in [1.37,1.56]$.
The fixed-cone isolation parameters are $R_{\rm iso} = 0.3$ and $E_{T}^{\rm max} = 3\,\GeV$.
Jets are reconstructed using the anti-$k_t$ jet algorithm~\cite{Cacciari:2008gp} with radius $R = 0.4$.
Jets are required to have transverse momentum $p_T^j \in [63,144]\,\GeV$ and absolute pseudorapidity $|\eta^j| < 2.1$, and to be back-to-back with the isolated photon, $|\Delta \varphi| > 7\pi/8$.
All photon--jet pairs, with the identified hadron inside the jet, are retained in the analysis, but the cuts mainly select events with a single jet.

As in the semi-inclusive case, we set $\mu_r = \mu_f = p_T^\gamma$.
However, when hadrons are identified within jets, the finite jet radius introduces an intrinsic cutoff on extra radiation, effectively reducing the fragmentation scale~\cite{Dokshitzer:1987nm,Dokshitzer:1991wu,Procura:2009vm,Jain:2011xz,Kaufmann:2015hma}.
Therefore, we employ $\mu_a = R \cdot p_T^{j(h)}$.

\begin{figure}[ht]
  \centering
  \includegraphics[width=\columnwidth]{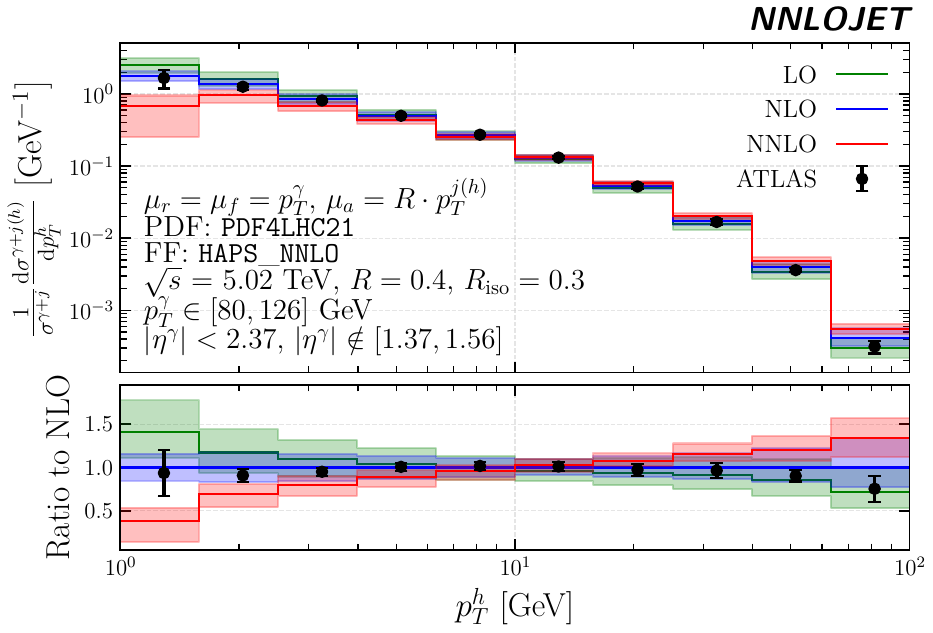}
  \caption{Data--theory comparison for $h^\pm$-in-jet production in association with an isolated photon up to NNLO with \HAPS.}
  \label{fig:atlas-absolute-pt}
\end{figure}
In Fig.~\ref{fig:atlas-absolute-pt} we compare our theoretical predictions up to NNLO for the hadron-in-jet cross section differential in the transverse momentum of the identified hadron $p_T^h$ to the ATLAS data.
The NLO distribution already provides a good description of the data, while the NNLO prediction, despite exhibiting reduced scale-variation uncertainties for larger values of $p_T^h$, shows tension with the experimental measurement.
In the low-$p_T^h$ region, i.e., for $p_T^h \in [1,4]\,\GeV$, the perturbative convergence is poorer, since we approach the small-$z$ regime, where resummation is required for reliable predictions.

\begin{figure}[ht]
  \centering
  \includegraphics[width=\columnwidth]{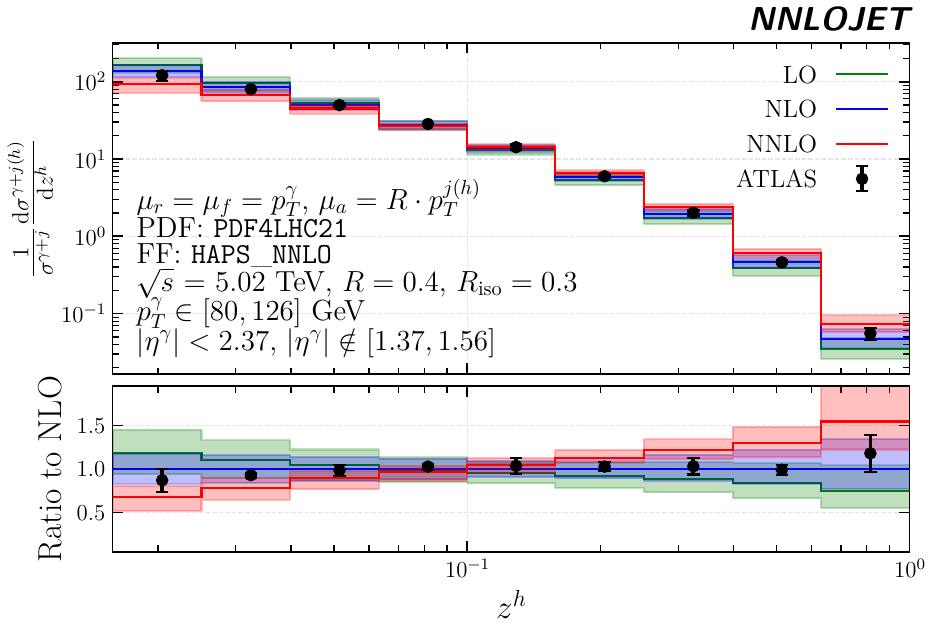}
  \caption{Data--theory comparison for $h^\pm$-in-jet production in association with an isolated photon up to NNLO with \HAPS.}
  \label{fig:atlas-absolute-z}
\end{figure}
In Fig.~\ref{fig:atlas-absolute-z} we compare our predictions to experimental data for the differential distribution in $z^h$, as defined in Eq.~\eqref{eq:zh}.
Analogously to Fig.~\ref{fig:atlas-absolute-pt}, the NLO distribution provides the best description of the measurement, while the NNLO curve lies outside the NLO uncertainty band.
However, since events with harder identified hadrons also contribute to the small-$z^h$ regime, the perturbative convergence slightly improves in this region compared to the small-$p_T^h$ region shown in Fig.~\ref{fig:atlas-absolute-pt}.

\begin{figure}[ht]
  \centering
  \includegraphics[width=\columnwidth]{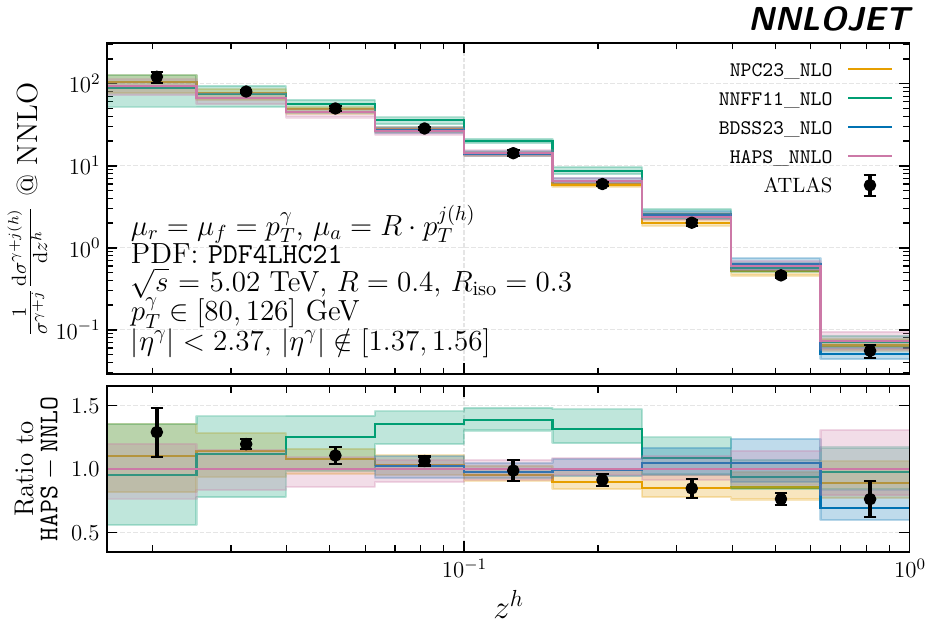}
  \caption{Data--theory comparison for $h^\pm$-in-jet production in association with an isolated photon at NNLO with different FF sets.}
  \label{fig:atlas-comparison-z}
\end{figure}
In Fig.~\ref{fig:atlas-comparison-z} the NNLO predictions produced with different FF sets are compared.
The prediction obtained with the \BDSS FF set provides a satisfactory description of the data within the
range of validity of the set, i.e., for the FF convolution variable $z \in [0.05,1]$.
For $z^h \in [0.06,0.25]$ the prediction produced with the \NNFF FF set overshoots both the data and
the other predictions, similarly to the semi-inclusive hadron production case presented in Fig.~\ref{fig:alice-comparison}.
Finally, the predictions computed with the \NPCnlo FF set achieve an excellent description of the ATLAS dataset, although this dataset is
included in the NLO global fit of that FF set~\cite{Gao:2024dbv}.
Interestingly, the NLO predictions shown in Ref.~\cite{Gao:2024dbv} display a moderate tension with the data, which may be
alleviated once our NNLO corrections to the differential distribution are taken into account.

We have checked (at the NLO level) that the inclusion of FF uncertainties in
Fig.~\ref{fig:atlas-comparison-z}, and more generally in all the predictions
presented in this Letter, does not alter our conclusions, as they
are subleading compared to the uncertainty due to scale
variations.
Similarly, PDF uncertainties were studied at NLO and found to be negligible
with respect to other sources of theoretical uncertainty.
This can be attributed to the normalisation of the distribution, which renders the prediction
largely insensitive to PDF effects.
\newline

\textit{Conclusions---}\ignorespaces
In this Letter, we have presented the first NNLO QCD predictions for
photon-associated hadron production at the
LHC, and compared them to available experimental data from the ALICE and ATLAS
Collaborations.
We have found NNLO corrections to be significant and, in most cases, to modify
the shape of the predicted distributions, motivating their inclusion in the
context of phenomenological studies.

Furthermore, we have investigated the impact of different FF sets on our predictions,
finding substantial differences among the resulting predictions, especially in the case of
hadron-in-jet observables.
This motivates the inclusion of these photon-plus-hadron data in FF
fits, given their potential to constrain flavour combinations (including
the gluon FF) that are only loosely constrained by other processes.

In general, predictions at NNLO accuracy for hadron production in $pp$
collisions are a key missing ingredient for promoting global FF fits to the
NNLO level.
In global fits that already include LHC photon-plus-hadron data at NLO, NNLO predictions
may alleviate tensions between datasets, thus potentially reducing FF extraction uncertainties.

Finally, given the role played by photon-plus-hadron production in jet-quenching
analyses, our results will be essential in view of HL-LHC heavy-ion runs,
providing a precise prediction for the proton--proton baseline.
\newline

\textit{Acknowledgments---}\ignorespaces
We thank Leonardo Bonino for his many contributions to fragmentation-related processes in \NNLOJET and Thomas Gehrmann for valuable discussions and comments on the manuscript.
The work of GF, AG, and FM is supported by the Swiss National Science Foundation (SNF) under contract 200021-231259 and by the Swiss National Supercomputing Centre (CSCS) under project ID ETH5f.
The work of GS is supported by the EU Horizon Europe research and innovation programme under the Marie Sk\l{}odowska-Curie Action HIPFLAPP, grant agreement 101149076.


\bibliography{bibliography.bib}

\end{document}